\catcode`\@=11

\newif\iffirstchoice@
\firstchoice@true
\def\textfonti{\the\textfont\@ne}
\def\textfontii{\the\textfont\tw@}
\def\text{\RIfM@\expandafter\text@\else\expandafter\text@@\fi}
\def\text@@#1{\leavevmode\hbox{#1}}
\def\text@#1{\mathchoice
 {\hbox{\everymath{\displaystyle}\def\textfonti{\the\textfont\@ne}%
  \def\textfontii{\the\textfont\tw@}\textdef@@ T#1}}
 {\hbox{\firstchoice@false
  \everymath{\textstyle}\def\textfonti{\the\textfont\@ne}%
  \def\textfontii{\the\textfont\tw@}\textdef@@ T#1}}
 {\hbox{\firstchoice@false
  \everymath{\scriptstyle}\def\textfonti{\the\scriptfont\@ne}%
  \def\textfontii{\the\scriptfont\tw@}\textdef@@ S\rm#1}}
 {\hbox{\firstchoice@false
  \everymath{\scriptscriptstyle}\def\textfonti
  {\the\scriptscriptfont\@ne}%
  \def\textfontii{\the\scriptscriptfont\tw@}\textdef@@ s\rm#1}}}
\def\textdef@@#1{\textdef@#1\rm\textdef@#1\bf\textdef@#1\sl\textdef@#1\it}
\def\DN@{\def\next@}
\def\eat@#1{}
\def\textdef@#1#2{%
 \DN@{\csname\expandafter\eat@\string#2fam\endcsname}%
 \if S#1\edef#2{\the\scriptfont\next@\relax}%
 \else\if s#1\edef#2{\the\scriptscriptfont\next@\relax}%
 \else\edef#2{\the\textfont\next@\relax}\fi\fi}

\def\Let@{\relax\iffalse{\fi\let\\=\cr\iffalse}\fi}
\def\vspace@{\def\vspace##1{\crcr\noalign{\vskip##1\relax}}}
\def\multilimits@{\bgroup\vspace@\Let@
 \baselineskip\fontdimen10 \scriptfont\tw@
 \advance\baselineskip\fontdimen12 \scriptfont\tw@
 \lineskip\thr@@\fontdimen8 \scriptfont\thr@@
 \lineskiplimit\lineskip
 \vbox\bgroup\ialign\bgroup\hfil$\m@th\scriptstyle{##}$\hfil\crcr}
\def\Sb{_\multilimits@}
\def\endSb{\crcr\egroup\egroup\egroup}
\def\Sp{^\multilimits@}

\newdimen\ex@
\def\rightarrowfill@#1{$#1\m@th\mathord-\mkern-6mu\cleaders
 \hbox{$#1\mkern-2mu\mathord-\mkern-2mu$}\hfill
 \mkern-6mu\mathord\rightarrow$}
\def\leftarrowfill@#1{$#1\m@th\mathord\leftarrow\mkern-6mu\cleaders
 \hbox{$#1\mkern-2mu\mathord-\mkern-2mu$}\hfill\mkern-6mu\mathord-$}
\def\leftrightarrowfill@#1{$#1\m@th\mathord\leftarrow\mkern-6mu\cleaders
 \hbox{$#1\mkern-2mu\mathord-\mkern-2mu$}\hfill
 \mkern-6mu\mathord\rightarrow$}
\def\overrightarrow{\mathpalette\overrightarrow@}
\def\overrightarrow@#1#2{\vbox{\ialign{##\crcr\rightarrowfill@#1\crcr
 \noalign{\kern-\ex@\nointerlineskip}$\m@th\hfil#1#2\hfil$\crcr}}}

\def\overleftarrow{\mathpalette\overleftarrow@}
\def\overleftarrow@#1#2{\vbox{\ialign{##\crcr\leftarrowfill@#1\crcr
 \noalign{\kern-\ex@\nointerlineskip}$\m@th\hfil#1#2\hfil$\crcr}}}
\def\overleftrightarrow{\mathpalette\overleftrightarrow@}
\def\overleftrightarrow@#1#2{\vbox{\ialign{##\crcr\leftrightarrowfill@#1\crcr
 \noalign{\kern-\ex@\nointerlineskip}$\m@th\hfil#1#2\hfil$\crcr}}}
\def\underrightarrow{\mathpalette\underrightarrow@}
\def\underrightarrow@#1#2{\vtop{\ialign{##\crcr$\m@th\hfil#1#2\hfil$\crcr
 \noalign{\nointerlineskip}\rightarrowfill@#1\crcr}}}

\def\underleftarrow{\mathpalette\underleftarrow@}
\def\underleftarrow@#1#2{\vtop{\ialign{##\crcr$\m@th\hfil#1#2\hfil$\crcr
 \noalign{\nointerlineskip}\leftarrowfill@#1\crcr}}}
\def\underleftrightarrow{\mathpalette\underleftrightarrow@}
\def\underleftrightarrow@#1#2{\vtop{\ialign{##\crcr$\m@th\hfil#1#2\hfil$\crcr
 \noalign{\nointerlineskip}\leftrightarrowfill@#1\crcr}}}

\catcode`\@=\active

\def\frac#1#2{{#1 \over #2}}

\newcount\GRAPHICSTYPE
\def\GRAPHICSPS#1{%
\ifnum\GRAPHICSTYPE=1 language "PS", include "#1"\else%
ps: #1\fi}

\def\graffile#1#2#3#4{\leavevmode\raise -#4 \hbox{%
\raise #3 \hbox{\rule{0.003in}{0.003in}\special{#1}}}%
{\raise -#4 \hbox to #2 {\vrule height#3 width0in depth0in\hfil}}%
}

\def\draftbox#1#2#3#4{\leavevmode\raise -#4 \hbox{\frame{\rlap{\protect\tiny
#1}%
\hbox to #2{\vrule height#3 width0in depth0in\hfil}}}}

\newcount\draft
\def\GRAPHIC#1#2#3#4#5{\ifnum\draft=1 \draftbox{#2}{#3}{#4}{#5}\else%
\graffile{#1}{#3}{#4}{#5}\fi}

\def\addtoLaTeXparams#1{\edef\LaTeXparams{\LaTeXparams #1}}

\def\doFRAMEparams#1{\readFRAMEparams#1\end}
\def\readFRAMEparams#1{%
\ifx#1\end%
\let\next=\relax%
\else%
\ifx#1i%
\dispkind=0%
\fi%
\ifx#1d%
\dispkind=1%
\fi%
\ifx#1f%
\dispkind=2%
\fi%
\ifx#1t%
\addtoLaTeXparams{t}%
\fi%
\ifx#1b%
\addtoLaTeXparams{b}%
\fi%
\ifx#1p%
\addtoLaTeXparams{p}%
\fi%
\ifx#1h%
\addtoLaTeXparams{h}%
\fi%
\let\next=\readFRAMEparams%
\fi%
\next%
}

\def\IFRAME#1#2#3#4#5{\GRAPHIC{#5}{#4}{#1}{#2}{#3}}

\def\DFRAME#1#2#3#4{
  \begin{center}
    \GRAPHIC{#4}{#3}{#1}{#2}{0in}
  \end{center}
}

\def\FFRAME#1#2#3#4#5#6#7{
  \begin{figure}[#1]
    \begin{center}
      \GRAPHIC{#7}{#6}{#2}{#3}{0in}
    \end{center}
    \caption{\label{#5}#4}
  \end{figure}
}

\def\FRAME#1#2#3#4#5#6#7#8{%
\newcount\dispkind%
\def\LaTeXparams{}%
\dispkind=0%
\def\LaTeXparams{}%
\doFRAMEparams{#1}%
\ifnum\dispkind=0%
\IFRAME{#2}{#3}{#4}{#7}{#8}%
\else
  \ifnum\dispkind=1
    \DFRAME{#2}{#3}{#7}{#8}
  \else
    \ifnum\dispkind=2
      \FFRAME{\LaTeXparams}{#2}{#3}{#5}{#6}{#7}{#8}
    \fi
  \fi
\fi
}

\catcode`\@=11

\long\def\QQQ#1#2{}
\def\QTP#1{}
\long\def\QQA#1#2{}

\def\EXPAND#1[#2]#3{}
\def\NOEXPAND#1[#2]#3{}

\def\LaTeXparent#1{}

\@ifundefined{INDEX}{\def\INDEX#1#2{}{}}{}
\@ifundefined{SUBINDEX}{\def\SUBINDEX#1#2#3{}{}{}}{}
\def\initial#1{\bigbreak{\raggedright\large\bf #1}\kern 2pt\penalty3000}

\@ifundefined{abstract}{%
\def\abstract{\if@twocolumn
\section*{Abstract (Not appropriate in this style!)}
\else \small
\begin{center}
{\bf Abstract\vspace{-.5em}\vspace{0pt}}
\end{center}
\quotation
\fi}}{}
\@ifundefined{endabstract}{%
\def\endabstract{\if@twocolumn\else\endquotation\fi}}{}
\@ifundefined{maketitle}{\def\maketitle#1{}}{}
\@ifundefined{affiliation}{\def\affiliation#1{}}{}
\@ifundefined{proof}{}{}
\@ifundefined{newfield}{\def\newfield#1#2{}}{}
\@ifundefined{chapter}{\def\chapter#1{\par(Chapter head:)#1\par }}{}
\@ifundefined{part}{\def\part#1{\par(Part head:)#1\par }}{}
\@ifundefined{section}{\def\part#1{\par(Section head:)#1\par }}{}

\makeatletter

\documentstyle[12pt]{article}

\author{M. Temple-Raston\\
Department of Physics,\\
Bishop's University,\\
Lennoxville, Quebec CANADA}
\title{Solitons in topological field theories
}

\begin{document}

\maketitle
\begin{abstract}
We present a topological Lagrangian field theory that is geometrically
similar to Yang-Mills(-Higgs) theory, and study the Bogomol'nyi solitons
contained within this theory. The topological field theory may provide an
example of a dual field theory.  The existence of a dual field theory to
Yang-Mills(-Higgs) theory was conjectured by Montonen and Olive.
\end{abstract}

\section{Introduction}

Recently a class of Lagrangian topological field theories possessing a
`minimizing' Bogomol'nyi structure has been introduced on oriented, compact,
connected four-manifolds \cite{mtr}. The associated Bogomol'nyi equations
are reminiscent of the self-duality equations in Yang-Mills theory, and the
solutions to the topological Bogomol'nyi equations share much in common with
solutions to the self-duality equations (instantons). Like the Yang-Mills
instanton, for example, solutions to the topological Bogomol'nyi equations
can be translated into geometrical structure on an appropriate holomorphic
vector bundle, and, the moduli space of solutions forms a Hausdorff
differentiable manifold. We shall call solutions to the topological
Bogomol'nyi equations `topological instantons'. The topological field
theories studied in \cite{mtr} achieve these results with relatively little
hard analysis and algebraic geometry when compared with the Yang-Mills
instanton theory \cite{don}. The reason for this is that topological
instantons are essentially equivalent to the differential geometric
formulation of `stable vector bundles' due to Kobayashi \cite{kob}. The
differences between Yang-Mills instantons and topological instantons are
also significant. We mention three differences.  First, non-trivial
topological instantons can exist on pseudo-Riemannian space-times,
while Yang-Mills instantons are trivial on space-times.  Second,
topological instantons have a larger gauge group, $U(n)$. Third,
topological instantons by virtue of their non-triviality on space-times have
a space-of-motions equivalent to the moduli space; Yang-Mills instantons are
pseudo-particles and do not possess a space-of-motions. It is well-known
that self-dual instantons in Yang-Mills theory and BPS magnetic monopoles in
Yang-Mills-Higgs theory are closely related \cite{atiyah}. BPS magnetic
monopoles are non-singular, finite-energy solutions to the self-duality
equations reduced to three spatial dimensions with a gauge symmetry in the
(imaginary) time direction. A similar process can be applied to the
topological instanton, leading to the theory of topological monopoles. The
topological instanton and the topological monopole obtained by dimensional
reduction are the subject of this paper.

In the next section we discuss the differential geometry of the class of
topological field theories on four-manifolds introduced in \cite{mtr}, and
expose the Bogomol'nyi structure. Solutions to the Bogomol'nyi equations
(topological instantons) are shown to be projectively flat. The physical
stability of the topological instanton field configuration is argued from
the topology of the underlying four-manifold. In section three, we
dimensionally reduce the four-dimensional topological field theory to three
spatial dimensions. The Bogomol'nyi structure survives the dimensional
reduction. Topological monopoles are the solutions to the Bogomol'nyi
equations in three dimensions.
Although the theory of topological monopoles is very similar to
the theory of BPS magnetic monopoles, there is an interesting difference
between the Bogomol'nyi structures of the two theories. In the theory of BPS
magnetic monopoles the Bogomol'nyi equations appear as a completed square in
the Lagrangian, while in the theory of topological monopoles they do not.
The Bogomol'nyi equations in our class of TFTs consist of two equations,
either of which will saturate the Bogomol'nyi energy. This added flexibility
in saturating the Bogomol'nyi energy allows greater freedom in constructing
solitonic particles with either an electric or magnetic charge.

\section{Instantons in topological field theories}

The Lagrangian theories in \cite{mtr} are defined by the Lagrangian Action
functional:
\begin{equation}
\label{TFT4}{\cal L}(A,B)=\int_M<(H^A\otimes I_E)\wedge (I_E\otimes
K^B)>-\frac 12<(I_E\otimes K^B)^2>
\end{equation}
defined on the product space ${\cal A}(P)\times {\cal A}(P)$. Interpreting
$H^A$ and $K^B$ as curvatures in the Lagrangian Action requires that the
real dimension of $M$ be four. $I_E$ is the identity transformation on the
adjoint bundle, $E$. The brackets $<\ >$ remind us that a choice of
adjoint-invariant, real-valued inner product on the adjoint bundle is
needed. The Action functional introduces an artificial asymmetry in $H^A$
and $K^B$ which is not supported by a physical argument; we will return to
this later. The variational field equations for (1) are
\begin{equation}
\label{field}D^AK^B=0,\qquad D^BH^A=0,
\end{equation}
where we have made use of the Bianchi identity $D^BK^B=0$. The set of
solutions is clearly neither empty nor entirely trivial. The physical
stability of a class of nontrivial, nonsingular, finite-Action solutions to
the variational equations (3) can be demonstrated by a topological argument.
The Lagrangian (\ref{TFT4}) can be rewritten as
\begin{equation}
\label{top}2{\cal L}(A,B)=-\int_M<(H^A\otimes I_E-I_E\otimes
K^B)^2>+\int_M<(H^A\otimes I_E)^2>
\end{equation}
The inner product structure defines a Weyl polynomial of degree two. Let
$E_A $ and $E_B$ be the vector bundle $E$ equipped with either the
connection
$A$ or $B$, respectively. The first term in the Lagrangian ${\cal L}$ in
equation (\ref{top}) is a topological invariant for the tensor product
bundle $E_A\otimes E_B^{*}$. Recall that the curvature of $E_A\otimes
E_B^{*} $ is given by $\Omega _{E_A\otimes E_B^{*}}=H^A\otimes
I_E-I_E\otimes K^B$. The Bogomol'nyi equations,
\begin{equation}
\label{Bog}H^A\otimes I_E=I_E\otimes K^B,
\end{equation}
are therefore a vanishing curvature condition on the tensor product bundle
$E_A\otimes E_B^{*}$. Solutions to (\ref{Bog}) automatically satisfy the
variational field equations (\ref{field}). An indice computation for (\ref
{Bog}), $H_{ab}^A\ \delta _{cd}=\delta _{ab}\ K_{cd}^B$, shows that the
curvature forms $H^A$ and $K^B$ are projectively flat. That is,
\begin{equation}
\label{Bog*}H^A=K^B=iFI_r,
\end{equation}
where $F$ is a real-valued two form on $M$, and $I_r$ is the identity
endomorphism for the vector bundle, $E$, of rank $r$. The Bianchi identity
imposes a simple condition on $F$, that $dF=0$, so that
$F\in H^2(M,{\bf R})$. Since $M$ is compact, $H^2(M,{\bf R})$ is of
finite dimension. If $F$ is a
curvature on $M$, then the second term in (\ref{top}) is a topological
invariant of the underlying four-manifold, $M$. Topologically non-trivial
solutions to the Bogomol'nyi equations will be said to be `physically
stable' if $F$ is a curvature of $M$ and if the solutions have a fixed
non-zero Action given by
$$
2{\cal L}=-\int_MF\wedge F=-24\pi \ {\rm sgn}(M)\ \ne 0,
$$
where the topological signature of the manifold, $M$, is denoted by ${\rm sgn
}(M)$. Physically stable, non-trivial solutions to the Bogomol'nyi equations
(\ref{Bog*}) on the vector bundle $(E,<\ >)$ are called topological
instantons \cite{mtr}.

\section{Monopoles in topological field theories}

We now examine static, non-singular solutions to the Bogomol'nyi equations
(\ref{Bog*}). By assuming a gauge symmetry in the direction of time, $X_t$,
we can dimensionally reduce the four-dimensional theory on ${\bf R}^4$
defined by (\ref{TFT4}), to a theory on ${\bf R}^3$. The reductions are
performed using the gauge symmetry equations,
\begin{equation}
\label{reduc}
\begin{array}{c}
H^A(X_t,\cdot )=-D^A\Phi _A, \\
K^B(X_t,\cdot )=-D^B\Phi _B.
\end{array}
\end{equation}
Dimensional reduction introduces the equivariant Lie algebra valued fields,
$\Phi _A,\Phi _B\in \Lambda ^0(R^3,{\rm End}(E))$, defined by $\Phi
_A=A(X_t)\ $ and $\Phi _B=B(X_t)\ $ \cite{manton}. We can either reduce the
Bogomol'nyi field equations (\ref{Bog*}) directly, or, reduce the full
variational field equations. In the first case, the Bogmol'nyi equations
locally reduce to
\begin{equation}
\label{rBog}
\begin{array}{c}
D^A\Phi _A=D^B\Phi _B=EI_E, \\
H^A\left| _{{\bf R}^3}\right. =K^B\left| _{{\bf R}^3}\right. =FI_E.
\end{array}
\end{equation}
$E$ is the one-form obtained by contracting $F$ in (\ref{Bog*}) on the
infinitesimal time displacement. In the second equation in (\ref{rBog}) $F$
denotes the restriction of $F$ in four-dimensions restricted to the leaves
in the foliation defined by $X_t$. Alternatively, the Lagrangian Action (\ref
{TFT4}) after reduction becomes
\begin{equation}
\label{TFT3}
\begin{array}{rrl}
{\cal E}(A,B)=\int_{M_3}<(I_E\otimes K^B)\wedge (I_E\otimes D^B\Phi _B)> & -
& <(I_E\otimes K^B)\wedge (D^A\Phi _A\otimes I_E)> \\
-<(H^A\otimes I_E) & \wedge & (I_E\otimes D^B\Phi _B)>,
\end{array}
\end{equation}
and the dimensionally reduced field equations become
\begin{equation}
\label{field3}
\begin{array}{c}
D^BH^A=0,\qquad \qquad D^BD^A\Phi _A=\left[ H^A,\Phi _B\right] , \\
D^AK^B=0,\qquad \qquad D^AD^B\Phi _B=[K^B,\Phi _A].
\end{array}
\end{equation}
The energy functional (\ref{TFT3}) can be rewritten as
\begin{equation}
\label{top3}
\begin{array}{rll}
{\cal E}(A,B)=\int_{M_3}<(H^A\otimes I_E-I_E\otimes K^B) & \wedge & (D^A\Phi
_A\otimes I_E-I_E\otimes D^B\Phi _B)> \\
-\int_{M_3}<(D^A\Phi _A\otimes I_E) & \wedge & (H^A\otimes I_E)>
\end{array}
\end{equation}
It is clear from (\ref{top3}) that the reduced topological instanton
equations (\ref{rBog}) continue to saturate the Bogomol'nyi bound, given by
the second integral in (\ref{top3}). In Yang-Mills theory, solutions to the
time-reduced instanton equations are called BPS magnetic monopoles. We call
solutions to the time-reduced topological instanton equations: topological
monopoles. Unlike Yang-Mills theory, however, the energy functional (\ref
{top3}) is saturated at the Bogomol'nyi bound with either equation in (\ref
{rBog}). We need not insist that both equations in (\ref{rBog}) be satisfied
in order to saturate the bound, although of course the field configurations
must still satisfy the second-order variational field equations.

To be observable to conventional detectors, $U(n)$ field configurations must
be broken. The symmetry breaking mechanism for BPS magnetic monopoles is
very attractive \cite{Goddard}, so we will use it here. Imagine a solitonic
core region at the origin. Let $G$ and $H$ be compact and connected gauge
groups, where the group $H$ is assumed to be embedded in $G$. The gauge
group of the core region $G$ is spontaneously broken to $H$ outside of the
core region when the Higgs field is covariantly constant, $D\Phi =0$. In
regions far from the core ($r\rightarrow \infty $) where we assume that
$D^A\Phi _A=0$, it can be shown that
\begin{equation}
\label{s.s.b.}H^A=\Phi _AF_A,
\end{equation}
where $F_A\in \Lambda ^2(M_3,E_H)$, is any closed two-form on $M_3$ taking
values in the $H$-Lie algebra bundle, denoted by $E_H$ here. A similar
expression to (\ref{s.s.b.}), $K^B=\Phi _BF_B$, can be written when $D^B\Phi
_B=0$. We assume that $<$$\Phi _A\Phi _A>=1$ when $r>>1$ and where
spontaneous symmetry breaking has occurred. When $G=U(n)$ and $H=U(1)$, $F_A$
becomes a pure imaginary two-form on $M_3$. Consider the Bogomol'nyi
solitons defined by (\ref{rBog}). Solutions to (\ref{rBog}) have an energy
topologically fixed by
\begin{equation}
\label{bound}
\begin{array}{ccc}
{\cal E} & = & -\int_{M_3}<(D^A\Phi _A\otimes I_E)\wedge (H^A\otimes I_E)>
\\
& = & -\int_{M_3}d<\Phi _AH^A>=-\int_{S^2}<\Phi _AH^A>,
\end{array}
\end{equation}
where $S^2$ is a large sphere surrounding the monopole core and lying
completely in a region where $D^A\Phi _A=0$. Substituting (\ref{s.s.b.})
into (\ref{bound}) and using the normalisation condition $<$$\Phi _A\Phi
_A>=1$, the energy is fixed by $\int F_A$. As in the case of the BPS
magnetic monopole, $\int F_A$ would be interpreted as the magnetic charge.

\section{Conclusion}

In this short contribution we have introduced a class of topological field
theories in three and four dimensions, exposed their Bogomol'nyi structures,
and argued the physical stability of solutions. But we believe that the
theories presented here are incomplete because there is a physical asymmetry
in the gauge fields present in (\ref{TFT4}). Symmetry in the Action is
easily regained, however, by exchanging $H^A$ and $K^B$, and adding it to
the Lagrangian (\ref{TFT4}). The variational field equations and the
Bogomol'nyi equations are unchanged by the symmetrization. In four
dimensions, the stability of the topological instanton is only slightly
different---the symmetrized Action is twice that of the asymmetric Action.
In three dimensions, the saturated energy functional becomes
\begin{equation}
\label{gbog}{\cal E}=-\int_{M_3}<(D^A\Phi _A\otimes I_E)\wedge (H^A\otimes
I_E)>-\int_{M_3}<(D^B\Phi _B\otimes I_E)\wedge (K^B\otimes I_E)>.
\end{equation}
We argued stability from the topological interpretation that can be given to
(\ref{gbog}). The symmetrization of the topological field theory implies
that the solitonic particle is topologically stable if either of the
integrals in (\ref{gbog}) is non-vanishing. The integrals should correspond
to the magnetic and electric charge of the soliton given by $\int_{S^2}F_A$
and $\int_{S^2}F_B$, respectively.

A particularly glaring omission in the solitonic particle spectrum in YMH
theory is the electric monopole. The Montonen-Olive conjecture addresses
this by proposing with some compelling evidence that there exists a dual
field theory to YMH theory which would replace the BPS magnetic monopole
with solitonic intermediate vector bosons: $W^{\pm }$, $Z^0$ \cite{Olive}.
Although there is still much study needed, we believe that theory of
topological monopoles may be an example of a dual field theory \cite{mtr2}.
If so, then in order to ensure the stability of the $Z_0$ particle, the $Z_0$
must be a magnetic monopole.

\medskip\

\noindent Acknowledgements. We are grateful to NSERC for a research grant
supporting this work (OGP0105498).

\end{document}